\documentclass{article}

\usepackage[utf8]{inputenc}
\usepackage{graphicx}
\usepackage{amsmath}
\usepackage{authblk}
\usepackage{braket}
\usepackage{verbatim}

\usepackage{hyperref}

\usepackage{xcolor}
\definecolor{darkred}{rgb}{0.7, 0.0, 0.0}

\title{Time versus Hardware: Reducing Qubit Counts with a (Surface Code) Data Bus}

\author[1,2,*]{Daniel Herr}
\author[3]{Alexandru Paler}
\author[4]{Simon J. Devitt}
\author[1,5]{Franco Nori}

\affil[*]{daniel.herr@riken.jp}
\affil[1]{Theoretical Quantum Physics Laboratory, RIKEN Cluster for Pioneering Research, Saitama 351-0198, Japan}
\affil[2]{Computational Physics, ETH Zurich, 8093 Zurich, Switzerland}
\affil[3]{Linz Institute of Technology, Johannes Kepler University, Linz, 4040, Austria}
\affil[4]{Centre for Quantum Software \& Information (QSI), Faculty of Engineering \& Information Technology, University of Technology Sydney, Sydney, NSW 2007, Australia}
\affil[5]{Department of Physics, University of Michigan, Ann Arbor, MI 48109--1040, USA}

\date{February 2019}

\begin{document}

\maketitle

\begin{abstract}
We introduce a data bus, for reducing the qubit counts within quantum computations (protected by surface codes). For general computations, an automated trade-off analysis (software tool and source code are open sourced and available online) is performed to determine to what degree qubit counts are reduced by the data bus: is the time penalty worth the qubit count reductions? We provide two examples where the qubit counts are convincingly reduced: 1) interaction of two surface code patches on NISQ machines with 28 and 68 qubits, and 2) very large-scale circuits with a structure similar to state-of-the-art quantum chemistry circuits. The data bus has the potential to transform all layers of the quantum computing stack (e.g., as envisioned by Google, IBM, Riggeti, Intel), because it simplifies quantum computation layouts, hardware architectures and introduces lower qubits counts at the expense of a reasonable time penalty.
\end{abstract}

\section{Introduction}

Useful quantum algorithms are reliant on error correction~\cite{LB13,qec_book}. Current quantum computers do not have the required qubit capacity to implement useful algorithms with a sufficient degree of fault tolerance. There are still too few qubits, and, despite being below threshold (for some operations), the error rate is still high. Furthermore, the overhead for fault-tolerant computation is huge.

To allow useful quantum computation as soon as possible, algorithms need to be optimized such that less logical qubits are required. Additionally, fault-tolerant computation protocols need to be optimized such that the overhead of physical qubits is reduced. For the latter optimization, we turn our attention to the surface code~\cite{Fowler2012,Raussendorf2007,BK05} as the underlying error-correction protocol, and lattice surgery~\cite{Horsman2012} as a means of computation.

Recent works~\cite{improvement_litinski, improvement_fowler, herr_lattice_surgery} reduce the overhead in time and physical qubits by moving to lattice-surgery-based methods of computation. In particular,~\cite{improvement_litinski} achieves an order of magnitude improvement over braiding with the introduction of multi-qubit measurements. These measurements require all-to-all connectivity between logical qubits, and logical ancilla patches are introduced for this.

In this paper, \emph{we remove the space overhead of the ancillary patches}. We introduce a \emph{thin} data bus between planar-code patches instead of the ancilla patches (e.g., Fig.~\ref{fig:architecture}). We show that this approach is completely consistent with the surface code and can be implemented on any architecture that supports the surface code. However, nothing comes for free, and there is a trade-off to be made: while our scheme \emph{reduces the overhead for all ancilla patches and physical qubits}, the time to execute a multi-qubit measurement takes longer depending on the distance of the surface code.

Our work is motivated by a realistic observation: Considering the slow increase of the qubit counts in quantum hardware, any method that enables the reliable execution of relevant quantum computations is welcomed. The herein presented trade-off analysis is based on worst-case assumptions and approximations. We also link to an online tool which we authored to automatically analyze and visualize, in great detail, the trade-offs of different optimization heuristics within the surface code. The presented trade-off results seem beneficial in most cases. We give two positive examples, and calculate, for each example, the total qubit savings when the data bus is used.

\section{Results}

\begin{figure}[t!]
    \centering
    \includegraphics[width=0.5\linewidth]{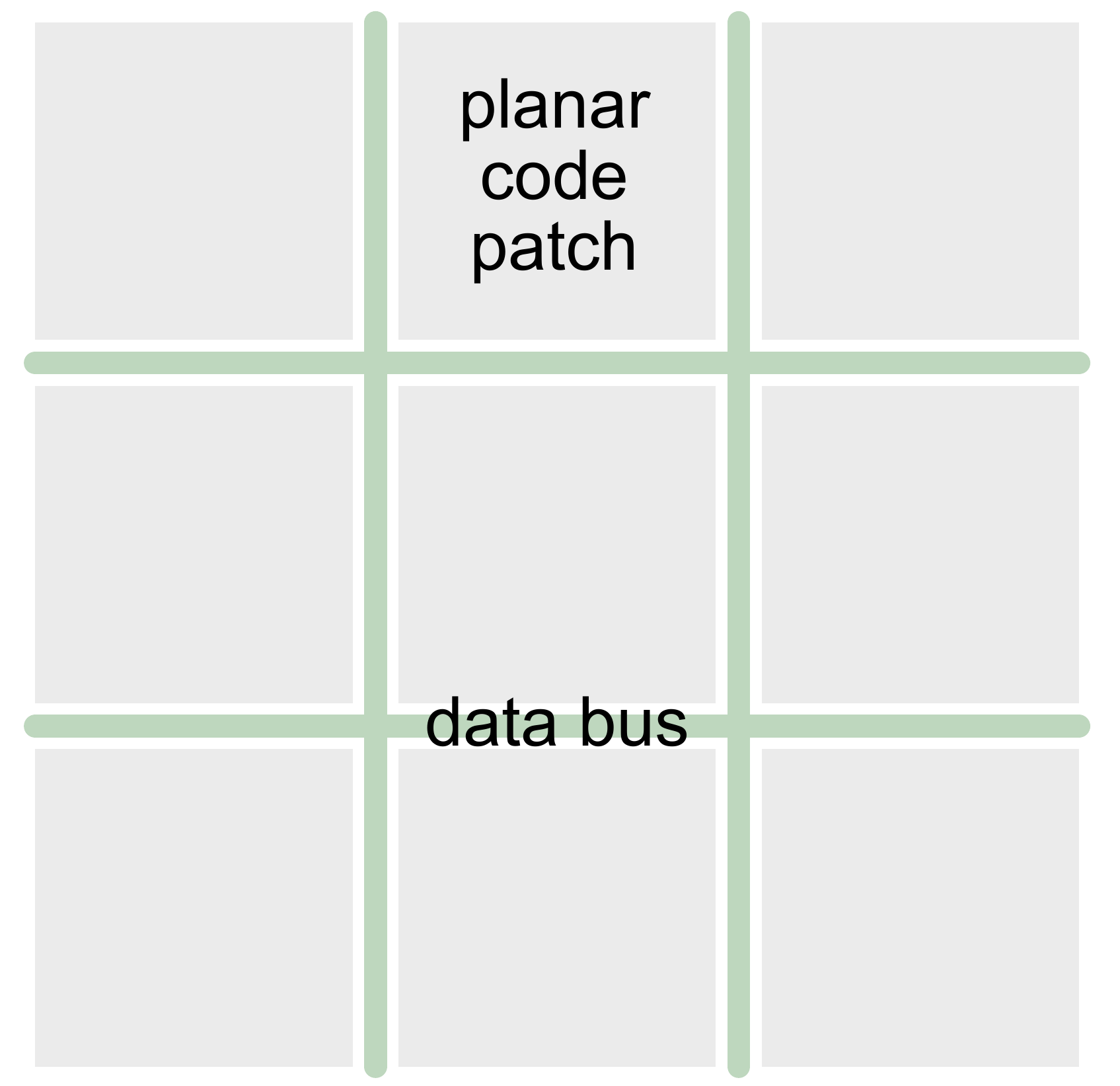}
    \caption{The grey squares are planar-code patches and the green lines make up the data bus. The data bus consists of a single layer of qubits that allows all-to-all connectivity between the data patches.}
    \label{fig:architecture}
\end{figure}

The basic idea of our approach is to \emph{use the physical qubits in between different patches of logical qubits as a data bus}. The data bus is a single long-range stabilizer~\cite{PhysRevLett.95.230504}, that acts on logical qubit states. Unlike surface-code stabilizers which are regularly measured during syndrome extraction, this logical stabilizer is only enforced once logical information has to be manipulated (i.e., merge, split operations). Information is manipulated by measuring logical qubit patches simultaneously. Thus, a single bit of information is necessary to interpret the measurement result, and the data bus is used to store this bit. This classical bit of information does not require the full error correction that a logical qubit needs. Instead, the bit can be encoded using simpler error correction methods such as the repetition code.

We propose the architecture shown in Figure~\ref{fig:architecture} (layout and architecture are used interchangeably in the following), where logical operators are stored on patches of planar code surrounded by a \emph{single string of repetition code}. The grid between different planar code patches has a thickness of only one (physical) qubit. This is in stark contrast to the thickness of ancilla patches where the thickness is given by the distance $d$ of the planar code.

The data bus implements the same operation as a merge operation does (i.e., a joint parity measurement). This parity measurement can be performed between arbitrarily many qubits. The measurement can be implemented fault-tolerantly (see next Section).

The data bus has several advantages. First, the overhead in terms of ancilla patches is reduced. Second, it allows arbitrary connectivity of different surface code patches without any movement. Third, $Y$-state measurements and parity checks are supported without any additional overhead during the computation.

The data bus supports both un-rotated \cite{Fowler2012} and rotated data patches \cite{Horsman2012}. In both the non-rotated and rotated layout of the planar code the qubits that are required during merge operations can be used to make up the data bus. The reduction in ancilla qubits does not come at the cost of adding additional qubits.

This data bus architecture can be used for communication between different quantum computers and allows interfaces between different fault-tolerant architectures. Lattice surgery between various codes has been shown to be possible in~\cite{interface_SSC}, and these methods can be extended to arbitrary connectivity using this data bus protocol. It might also be used as a means for quantum chip design~\cite{Wosnitzka2016}, where individual patches of surface code are placed far enough apart to minimize cross-talk. Similar to FPGA-design~\cite{FPGA} clusters of nearest neighbor patches that operate with higher speed are connected to each other by the data bus.

To see how many qubits can be saved by the data bus, we investigate two industrially relevant and practical scenarios:
\begin{enumerate}
    \item a simple noisy inter\-mediate-scale quantum (NISQ)~\cite{nisq} experiment that implements a joint parity check of arbitrary basis (e.g., $XX$, $XY$, $YZ$) on two logical qubits;
    \item quantum circuits where the Clifford part dominates the total of qubits necessary (a single distillation procedure is executed at a time, but the circuits operate on many qubits, e.g. 1024 qubit wide adder).
\end{enumerate}
In the NISQ case, a data bus architecture gives a $75\%$ reduction, from $157$ qubits to $55$ qubits, which makes it possible to implement the experiment on the current generation of quantum chips, for example the Bristlecone $72$-qubit chip~\cite{google_supremacy}. In the latter scenario, for large Clifford-dominated circuits, the data bus can provide reductions of roughly $15\%$ in the total number of qubits.

Hardware reductions come at the cost of time penalties: a classical trade-off. The disadvantage of the bus is that repetitions are necessary for computational fault-tolerance. Each preparation of the data bus requires $d$ rounds of stabilizer measurements. The whole procedure needs to be repeated $d$ times, and this results in a time cost per data bus usage of $d^2$. A merge operation only requires $d$ rounds of error correction. Thus, the time penalty of using the data bus, compared to not using it, is approximately $d$. The positive aspect is that the time penalty is not orders of magnitude, and its value is approximately the distance of the necessary surface code. All calculations assume a worst-case scenario for the data bus, where each operation is scaled by the distance $d$ in time. Nearest-neighbor interactions do not require the usage of the data bus and a hybrid model where the application of the data bus is mixed with traditional lattice surgery operations is very likely to lead to better results.

The factor $d$ time penalty increases the volume~\cite{volume_faulttolerant} of the fault-tolerant computation, which in turn is used to determine the surface code distance. As a result, in specific scenarios, the increased distance leads to larger qubit counts than when the data bus is not used. Nevertheless, the penalty is a worst-case estimation, and the discussion section sketches solutions (e.g., depth reduction by parallel quantum operations, data-bus sharing etc.) for reducing the time penalty and consequently qubit counts. The sketched solutions are typical computer engineering and architecture problems, and we are certain this work will spark the interest of those communities.

\begin{table}[t]
    \centering
    \begin{tabular}{l||r|r|r|r|r}
                          &\textbf{Q100} &\textbf{Chem 54} &\textbf{Chem 250} &\textbf{Shor 1024} &\textbf{Shor 4096}\\
        \hline\hline
        \textbf{Qubits (Q)}    & 100      & 123     & 341      & 3082      & 12298\\
        \textbf{Ancilla (A)}   & +50\%    & +50\%   & +50\%    & +50\%     & +50\%\\
        \textbf{Total (Q+A)}   & 150      & 185     & 512      & 4623      & 18447\\
        \hline
        \textbf{Volume}        & 1.31E+11 & 4.59E+9 & 7.56E+11 & 3.27E+14  & 8.37E+16 \\
        \textbf{Dist. w/o Bus} & 29       & 23      & 29       & 31        & 35\\
        \textbf{Dist. with Bus}& 31       & 25      & 31       & 35        & 39\\
        \hline
        \textbf{QC w/o Bus}    & 252300   & 195730  & 861184   & 8885406   & 45195150\\
        \textbf{QC with Bus}   & 204800   & 167648  & 700416   & 7988544   & 39353600\\
        \hline
        \textbf{QC Improv.}    & 0.81     & 0.86    & 0.81     & 0.90      & 0.87\\
        \hline
        \textbf{Hours w/o Bus} & 7.53      & 0.21    & 12.7     & n/a      & 49100\\
        \textbf{Hours with Bus}& 233.58    & 5.36    & 394.62   & n/a      & 1916243\\
        \hline\hline
    \end{tabular}
    \caption{The qubit counts, QC, of five circuits are compared.  Data bus usage influences the necessary distance, Dist, because of the time penalty. The improvements, QC Improv., range between $10\%$ and $18\%$. Here, w/o stands for without.}
    \label{tab:res}
\end{table}

Table~\ref{tab:res} includes qubit count (QC) estimations for quantum algorithms of practical interest: quantum chemistry \cite{babbush2018encoding} and Shor's algorithm \cite{Campbell2017}. We compare the qubit counts with the most up-to-date results from \cite{fowler_gidney_CCZ}. We verify the correctness of our count estimations by comparing the qubit counts without the data bus with the ones provided in \cite{fowler_gidney_CCZ}: the values are almost equal. The table shows that around $15\%$ qubit count reductions can be achieved by using the data bus (qubit counts with Bus). The time penalty is reflected in the hours of execution with data bus (Hours with Bus) compared to the original execution times (Hours without Bus).

The conclusion is that quantum chemistry simulations are still running in reasonable time, and require overall around $15\%$ less qubits.

For Shor's algorithm the time penalty is increasing an execution time which was not practical from the beginning: $49100$ hours are approximately five years, and the data bus increases the execution time (in the worst case) to $220$ years.

\section{Methods}

The data bus is a fault-tolerant long-range parity-check operation. Long-range parity checks on surface code architectures have been investigated before in~\cite{1704.02620,NSM16}. We use the same underlying method as in \cite{NSM16}, but for a different purpose. 

Long-range parity checks provide a recipe for performing arbitrary length fault-tolerant parity checks. We first describe the individual steps needed for the check operation. Afterwards we show that the operation has the desired properties.

\begin{enumerate}
\item Produce a GHZ-state with $N$ data qubits (four time steps), and keep it corrected against bit-flip errors ($d$ cycles).
\item Perform a transversal CNOT operation between all connected surface code patches and their part of the GHZ-state. The CNOTs can be applied in parallel and, thus, require only one time step.
\item Without the measurement of any further stabilizers, all data qubits of the complete chain of the GHZ-state are measured (one time step).
\item Repeat steps 1-3 for $d$ times, and use majority voting over the individual measurement results to obtain the error-corrected measurement result in total time $O(d^2)$.
\end{enumerate}

\subsection{Does the data bus have the intended effect?}

Yes. Let us take a look at the $XX$-parity check between logical qubits using the data bus. In the first step, a $N$-qubit GHZ-state has been prepared and its errors have been corrected. We can assume that the GHZ-state is given by:
\begin{equation}
	\ket{\psi} = \ket{00\ldots 0} + \ket{11\ldots 1}
\end{equation}
This GHZ-state can be expressed in the $X$-basis:
\begin{equation}
	\ket{\psi} = \frac{1}{\sqrt{N}}
    \sum_{\substack{s \in \left\{+,-\right\}^N\\
    		\text{even \# of } - \text{ in } s}}
            \ket{s}
\end{equation}
\begin{figure}[t]
    \centering
    \includegraphics[width=\linewidth]{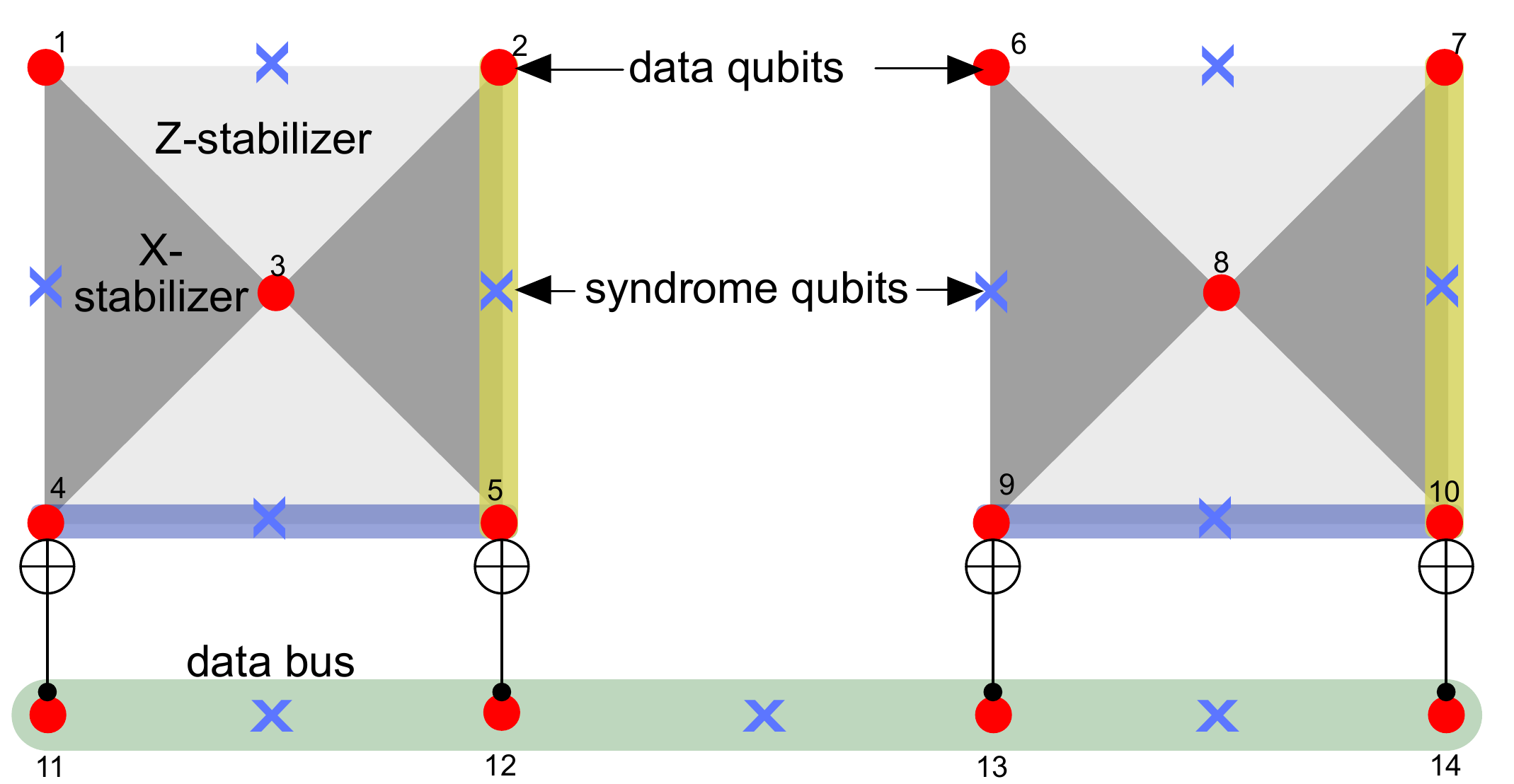}
    \caption{Example of a $XX$-parity check between two logical patches using the data bus. The illustrated surface code patches are of distance 2, but in practice arbitrary distances can be used. $X$-stabilizers are colored in dark grey, whereas $Z$-stabilizers are colored light grey. Data qubits of the planar code patches are red dots, and syndrome qubits are blue crosses. Logical $X$ and logical $Z$ operators are the blue and yellow chains. The data bus is colored in green, and the meaning of syndrome and data qubits are flipped (crosses are the data qubits of the repetition code). The CNOT operations are used between nearest-neighbor data qubits of the codes to perform a transversal CNOT.}
    \label{fig:XX}
\end{figure}
This means that the total parity of the GHZ-state is even in the $X$-basis. It also means that the total parity of this repetition code does not change under the application of any even number of physical $Z$-operations on the data qubits. Figure~\ref{fig:XX} shows how individual patches of logical qubits are connected to this data bus (GHZ-state) that supports $Z$-parity checks. The CNOTs are applied on a chain of qubits that correspond to the logical $X$-operator of the individual planar-code patches. Each of the patches will add an odd number of $Z$ operations to the GHZ-state if they are in the $\ket{-}$ state and thus flip the total parity. To flip the state from $\ket{+}$ to $\ket{-}$, a logical $Z$ operation needs to be applied. If a data qubit is in the $\ket{+}$ state, the total parity will not change. During the third step, a measurement in the $X$ basis of each individual qubit of the GHZ-state will result in the total parity of all logical qubits. The exact operation can also be seen on a stabilizer table, where the changes due to the CNOTs are colored in dark red.
\begin{equation*}
\begin{smallmatrix}
  1&2&3&4&5&6&7&8&9&10&11&12&13&14 \\
  X&&X&X& &&&& & & & & & \\
   & X&X& &X &&&& & & & & &\\
  Z&Z &Z&& &&&& & & & & &\\
  &&Z&Z &Z &&&& & & \textcolor{darkred}{Z}& \textcolor{darkred}{Z}& & \\
  &&&& &X&&X&X & & & & &\\
  &&&& &&X&X& &X & & & &\\
  &&&& &Z&Z&Z& & & & & &\\
  &&&& &&&Z&Z &Z & & &\textcolor{darkred}{Z} &\textcolor{darkred}{Z}\\
  &&&\textcolor{darkred}{X}&\textcolor{darkred}{X} &&&&\textcolor{darkred}{X}&\textcolor{darkred}{X} &X & X&X &X\\
  &&&& &&&&& &Z &Z & &\\
  &&&& &&&& & & & Z&Z &\\
  &&&& &&&& & & & &Z &Z\\
\end{smallmatrix}
\end{equation*}
The row with the $\textcolor{darkred}{X}$ operators is of particularly interest, because it represents a product of the total parity of the GHZ-state in the $X$-basis with the $XX$ parity of both patches. The total parity of this stabilizer is even. Therefore, one is able to deduce the logical $XX$ measurement with a measurement of the parity of the GHZ-state.

For a $ZZ$-parity check, the same general procedure can be performed, with the exception of the GHZ-state being initialized in the $X$-basis: $\ket{\phi} = \ket{+\ldots +} + \ket{-\ldots -}$.
The CNOTs need to connect the GHZ-state with the logical $Z$ operator chain and the CNOTs are inverted (target qubits of the CNOTs on the data bus). The final measurement needs to be in the $Z$-basis for the $ZZ$-parity check.

\subsection{Do arbitrary length parity-check operators accumulate errors?}

\begin{figure}[t]
    \centering
    \includegraphics[width=0.4\linewidth]{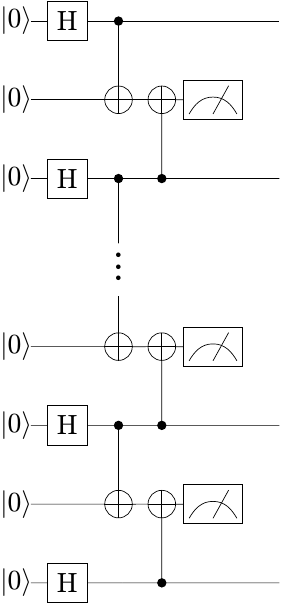}
    \caption{Generation of a GHZ-state. Every second qubit is a syndrome qubit whose measurement has to be repeated $d$ times.}
    \label{fig:gen_GHZ}
\end{figure}

No, and we demonstrate this by using the construction of arbitrary-size super operators as shown in Ref.~\cite{NSM16}. We review the construction for the $ZZ$ parity-check as was presented above. The main ingredient is the error-corrected $N$-qubit GHZ-state. The circuit in Figure~\ref{fig:gen_GHZ} creates the GHZ-state along a line of qubits where each second qubit is a syndrome qubit that can detect a change in the $ZZ$-parity of its neighbors. Each stabilizer measurement is repeated $d$ times to correct for measurement errors.

After the GHZ-state has been created and verified, a transversal CNOT between the surface-code patches and the GHZ-state can be performed. Immediately afterwards, the data qubits of the GHZ-state are measured.
Because this measurement is not protected, it has to be repeated $d$ times. The measurement will collapse the state into one of two sub-spaces. One subspace has an even number of Pauli-$X$ errors. The protected quantity of interest is the total parity along the data bus. An even number of errors does not influence the total result. The probability of this happening is given by
\begin{equation*}
	P_\text{even} = \mathcal{N} \sum_{i = 0}^{\infty} p_\text{phys}^{2i}
\end{equation*}
where $\mathcal{N}$ is a normalization factor to ensure that the probabilities of all odd and even outcomes sum to 1.

The other sub space has an odd number of errors and results in the wrong logical result. The probability for such an event is given by:
\begin{equation*}
	P_\text{odd} = \mathcal{N} \sum_{i=0}^{\infty} p_\text{phys}^{2i+1} =  \mathcal{N} p_\text{phys} \sum_{i=0}^{\infty} p_\text{phys}^{2i} = p_\text{phys} P_\text{even}
\end{equation*}
The odd subspace which results in a wrong measurement is less likely by a factor of $p_\text{phys}$ and repeated measurements can be performed for better protection against errors. For a $N$-qubit GHZ-state, $N$ measurements are needed. The majority of the measurement results will give the error-protected result of the parity between the two patches.

\subsection{How high is the time overhead?}

It is $d$, because we can show that even though the data bus consists of an $N$-qubit GHZ state, it needs only $d$ cycles of verification. The whole verification procedure is repeated only $d$ times. Thus, the overhead of the data bus is constant in its length and only depends on the distance $d$ of the surface code patches.

The success probability for a single logical qubit in unit time $P_L$ can be calculated using a fixed total success probability for the whole quantum algorithm, which is usually fixed as $P_\text{tot}$.
\begin{equation}
    P_{\text{tot}} = \left(1-P_L\right)^{V}
    \label{eq:global_prob}
\end{equation}
The volume $V$ is given by the product of the total number of logical qubits and the execution time for the whole circuit. This is a worst-case approximation because ancilla qubits are not active at all times and errors cannot affect the computation while the logical ancilla qubits are inactive.  For an arbitrary calculation, the ancilla qubits for routing can be replaced by the data bus. The data bus roughly introduces the same number of distance-$d$ corrected GHZ-states as there were ancilla patches. For a distance calculation, the number of logical qubits (or bits with the inclusion of the data bus) will not be changed, but the data bus will reduce the total number of physical qubits because the ancilla patches are now $d\cdot1$ instead of $d\cdot d$. However, the distance might increase while using the data bus because the volume increases (larger time).

The repetition code can be seen as a 1D version of the surface code (i.e., a single slice from the 2D surface)~\cite{fowler_repetition_surface}. Due to the restriction to one dimension, it can only correct for one type of errors. Nevertheless, the same decoding techniques (minimum-weight perfect matching) can be applied resulting in a comparable threshold and similar logical error rates~\cite{fowler_repetition_surface}. The distance $d$ has been chosen so that with probability $P_\text{tot}$ no error-chains of size $d/2$ appear at any logical qubit (data bus or otherwise). Thus, it suffices to use $d$ repetitions of syndrome measurements instead of $N$ to verify the GHZ state for a protection of up to $d/2$-qubit error chains.

The measurement procedure only needs to be of distance $d$ as well. This can be achieved by repeating the parity measurement at least $d$ times. The data bus is therefore a distance $d$ repetition code which is composed of $N$ qubits.

\subsection{Computation-specific analysis}

The applicability of the data bus is the result of a quantum algorithm and quantum circuit specific trade-off analysis. Such a trade-off analysis is performed by a software tool that allows to compare the footprint in total number of physical qubits for a variety of optimization heuristics. The tool is open-source and can be accessed at~\cite{link_open_source}.

The applicability of the data bus is presented in a NISQ experiment proposal, and in the case of large quantum circuits. The trade-off analysis is computation-specific, and should be executed for any circuit considered for optimization. The previous two examples benefit from the bus, but we present a counter-example too, which is the 15-to-1 $T$-state distillation circuit from \cite{improvement_litinski}.

In the following, we will refer to the \emph{safety factor}\cite{improvement_fowler}, which is an integer value related to the failure rate of an entire fault-tolerant quantum computation (total success probability for the quantum algorithm). To compare our qubit counts (resource estimations) with \cite{improvement_fowler} and \cite{fowler_gidney_CCZ} we also consider this factor.

Another parameter is the \emph{routing factor}, which is the ratio between the number of ancilla patches and the number of data patches. For example, in Table~\ref{tab:res} the routing factor is $0.5$, because $A = 0.5 \times Q$ across all circuits considered. A value of $0.5$ is both reasonable and practical in the presence of computer-generated layouts \cite{PDF16, paler2017synthesis, paler2019surfbraid, javadi2017optimized}.

\subsubsection{NISQ experiment}

To the best of our knowledge, surface code correction has not been demonstrated for more than one logical qubit. The following sketches the structure of a two-qubit experiment which, in theory, could be mapped to the Bristlecone 72-qubit machine~\cite{google_supremacy}.

In this setup, two logical qubits are encoded in surface code patches of distance two. This distance is too short, and only allows error detection but not correction. Post-selection should be used on the measurement results. Fault-tolerant measurement of any two qubit Pauli stabilizers (e.g., $XX$, $XY$, $ZY$) are performed.
\begin{figure}[t]
    \centering
    \includegraphics[width=\linewidth]{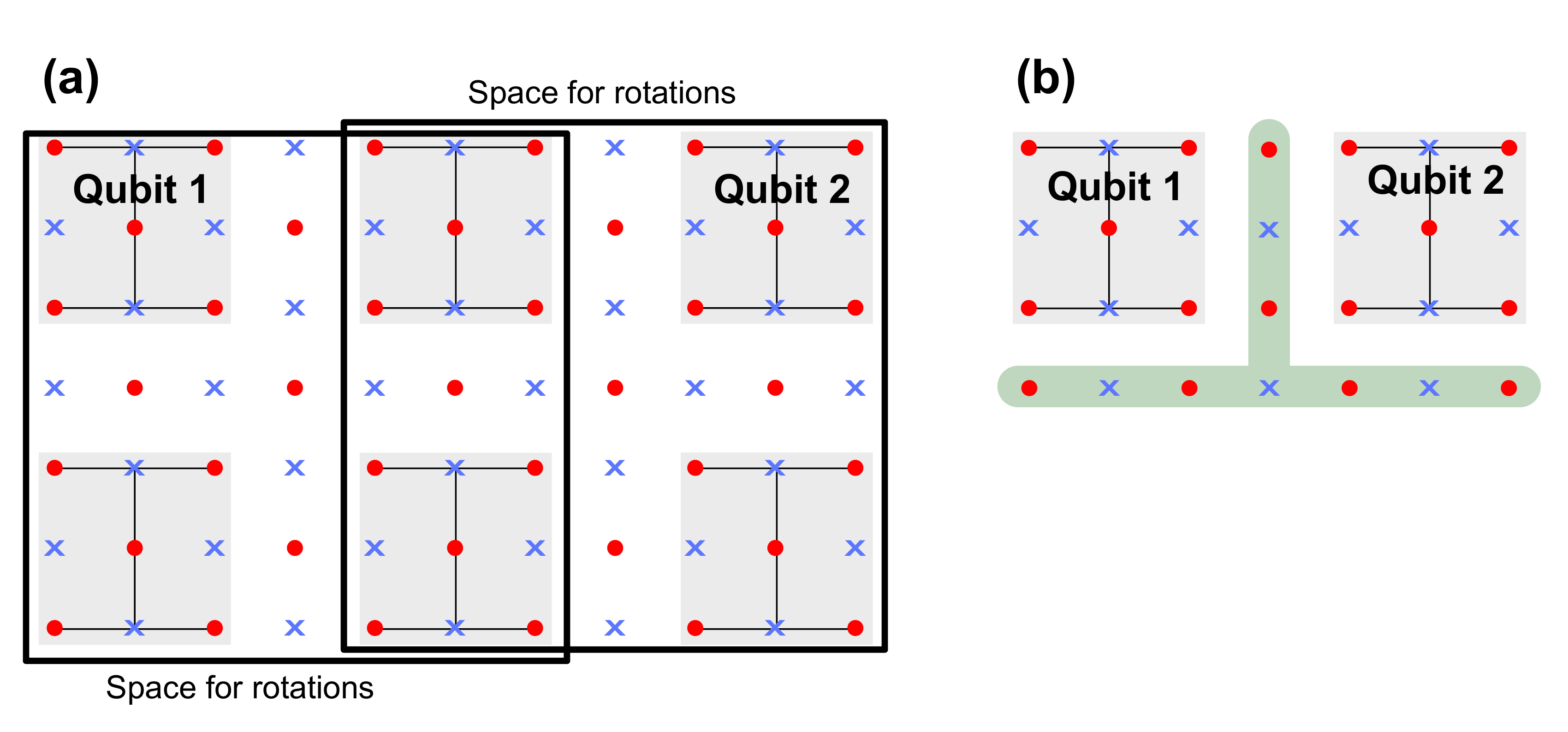}
    \caption{The smallest viable setup for arbitrary two-qubit parity checks. In (a) the setup with ancillas is shown. The additional ancilla patches are needed for rotation of the qubits and their connection. In (b) the data bus can perform the same operation using far less physical qubits.}
    \label{fig:NISQ}
\end{figure}

The qubit layouts for NISQ-style experiments: (a) without the data bus architecture, and (b) with the bus are given in Figure~\ref{fig:NISQ}. For the original layout, the total qubit count is $77$, and by using the data bus the qubit count is reduced to $28$. This assumes a realistic data bus size and not worst-case-bounded as in the Appendix.
The code distance of the experiment can be increased by using the data bus. For a rotated lattice with distance three and without the data bus the total qubit count is $151$. The data bus for distance-three-rotated patches results in a qubit count of $55$ physical qubits (it fits on Bristlecone).

The trade-off analysis tool~\cite{link_open_source} is programmed for rotated patches, and can be used for distances larger than two. We used this tool to determine the safety factors when using the data bus, and it ($487$; larger is better) is comparable  to the expected value ($99$).
\begin{figure}[t]
    \centering
    \includegraphics[width=0.7\linewidth]{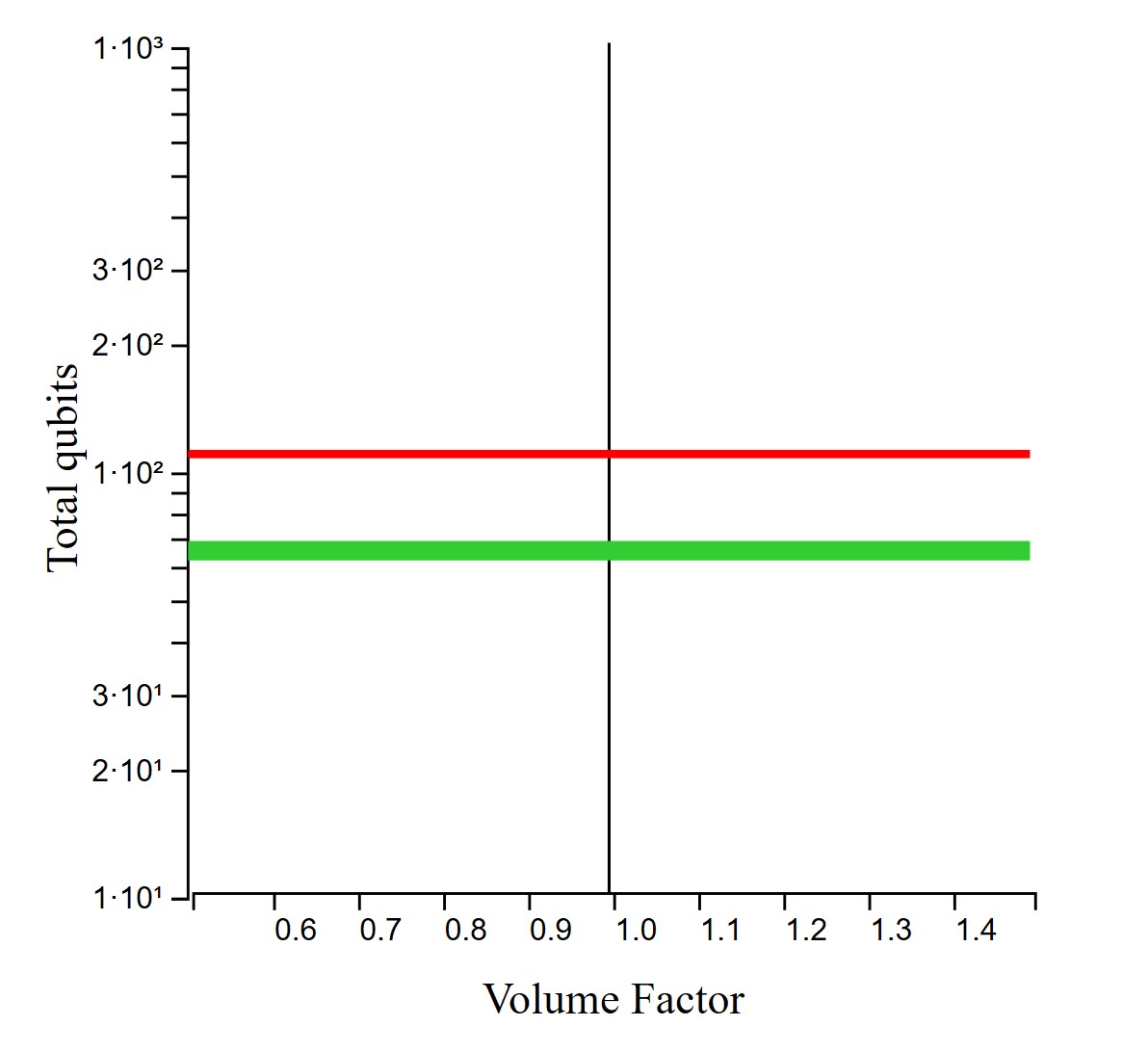}
    \caption{Similar circuits (the same spacial footprint, but different volumes) are compared. The horizontal axis indicates by which factor the volume of the circuit is changed. The vertical axis illustrates the total number of physical qubits for the original layout (red) and for the data bus layout (green). For $1.0$ on the horizontal axis, the volume is not scaled: it is the initial volume (the one which could be computed for example for surface code protected circuits).  In this plot, the distance is set to three and the safety factor is equivalent between the two volumes; nevertheless, the safety factor is always larger than a parameter specified by the designed circuit (e.g., 1\% failure rate of the entire computation).}
    \label{fig:NISQ_plot}
\end{figure}

\subsubsection{Large-scale computation}
\begin{figure}[t]
    \centering
    \includegraphics[width=0.7\linewidth]{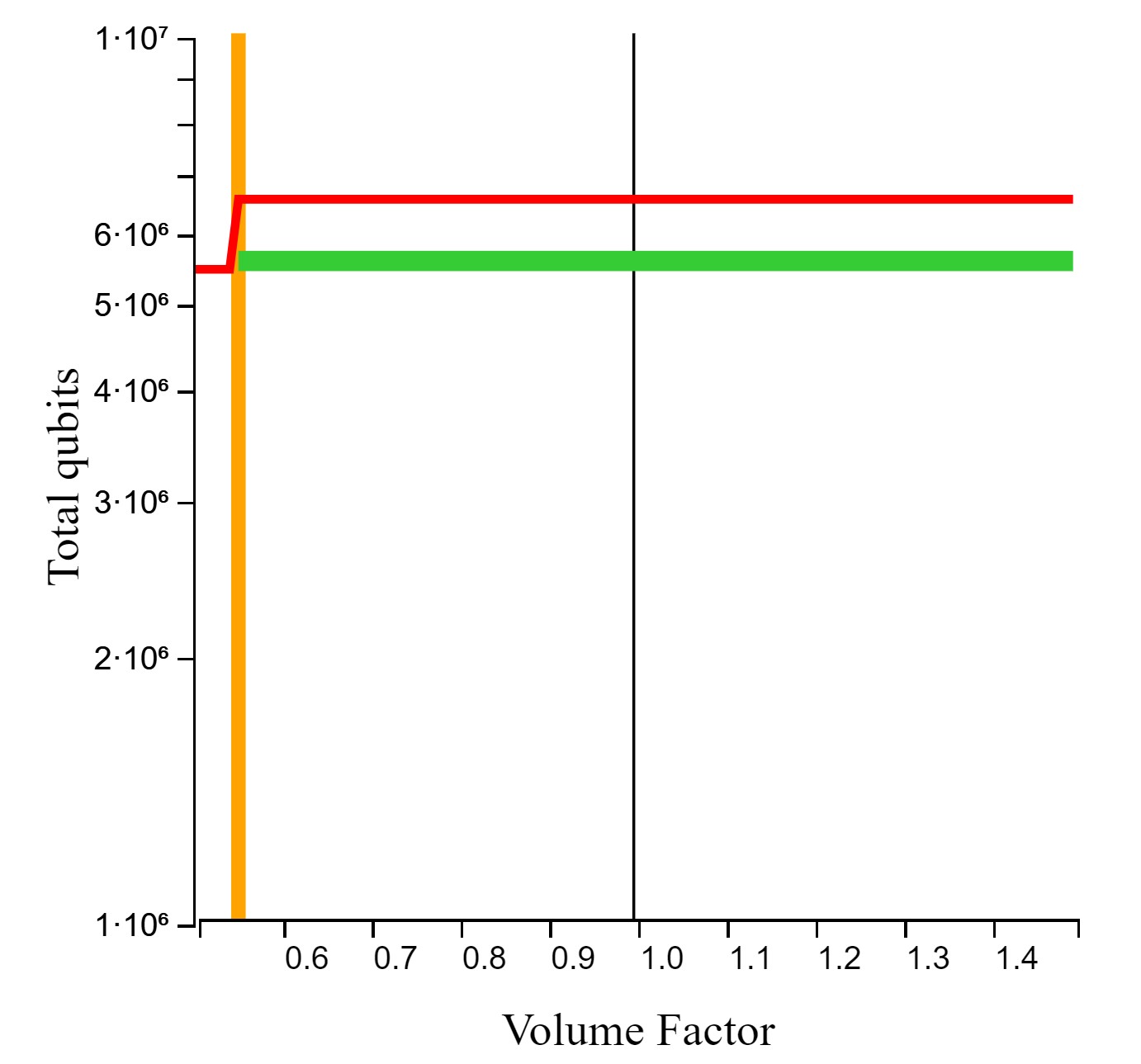}
    \caption{A plot similar to Figure~\ref{fig:NISQ_plot}. This plot compares the achieved qubit savings in the case of a large scale computation. The number of qubits required for the original architecture (red) is larger compared to when the data bus is used. Just before distance jumps, due to a fixed total success probability, the data bus is worse due to its increase in execution time.}
    \label{fig:large_scale}
\end{figure}

For large-scale computation, the routing often comprises one third of the number of qubits for a circuit~\cite{fowler_gidney_CCZ}. This entails routing the magic states from their distillation~\cite{BK05,jones2013low} to the qubits where they are injected. For large volumes, the application of a data bus seems to be beneficial, as can be seen in Figure~\ref{fig:large_scale}. An interactive version of this Figure is hosted at \cite{link_open_source}. In general, after a distance change, the addition of the data bus (green) is beneficial because the additional penalty in time does not necessitate another change in distance. 

The parameters considered for the plot from Fig.~\ref{fig:large_scale} were randomly chosen, but similar plots exhibit the same properties. This fact can be verified by using the online tool available at \cite{link_open_source} and by comparing the table from the Results section with the qubit count estimations from \cite{fowler_gidney_CCZ}. In particular, for Fig.~\ref{fig:large_scale}, the parameters are: volume is $6144\times 3000$, the number of patches (data and ancilla) is $6144$, the physical error rate is $0.1\%$. The qubit count is reduced about $15\%$, from $6,500,352$ (without data bus) to $5,537,792$ (with data bus). The reduction of the total qubits number can be attributed to fewer qubit requirements thanks to the data bus. A $4\times d^2$ patch is replaced by a $2d$ chain (including measurement qubits).

\subsection{Counter-example}

The data bus and the associated trade-off analysis were used to understand the advantages when applied to smaller volume circuits, such as $T$ state distillations. For this we used the distillation circuit presented in \cite{improvement_litinski}. That circuit has a volume of $11 \times 23$, where four patches are used for routing and the total number of patches is $11$. Thus, the routing factor is $7/11 \approx 0.63$. The qubit count did not reduce for any distance in the integer range $[15, 45]$. The reason why this circuit seems, at least for the moment, not automatically optimizable is that: a) it requires large distances, because they are used for suppressing injected errors; b) have low volumes, and are (semi-) hand optimized.

Other already hand-optimized circuits, such as those from \cite{fowler_gidney_CCZ}, are almost impossible to automatically optimize by using the data bus directly. Such circuits include patch connections and placements, which cannot be abstracted through a single parameter (e.g., the routing factor).

\section{Discussion}

The data bus is, from a technical perspective, a surface code patch protected against a single type of errors (instead of the two, bit and phase errors). This has the advantage that it reduces the qubit counts necessary for the ancillary patches used to intermediate between the logical qubit patches. The disadvantage is that hardware reductions come at the cost of time penalties. Nevertheless, considering the challenges of increasing the number of qubits, it may be worth to take this penalty in the first and maybe the second generation of quantum computers.

Intuition would suggest that removing the error protection on certain patches will reduce the fault-tolerance of the entire computation beyond control. Surprisingly, this work shows that this is not always the case. The successful usage of the data bus depends on the distance and on the depth of the volume estimating the computation (see Appendix for the intuition behind the plots). The cases for which the data bus reduces the qubit counts are the space-time volumes for which even with additional time overhead the same distance is sufficient for the quantum computation. The opposite cases are when the time penalty increases the total space-time volume, such that higher distances (and qubit counts) are used.


Instead of minimizing the number of logical qubits a quantum circuit operates on, in the presence of the data bus, it makes sense to parallelize as many of the single qubit gates. While the bus is operated, many data patches are, in the worst case considered herein, not used for computation.
This leaves, for each bus operation, a window of $d$-time length which could be used for pre-caching computational results to be communicated later along the bus (slowly for the moment, compared to individual data patches). This data bus model shows the possibility of directly using computer engineering \cite{hennessy2011computer,AnalogCircuitDesign}, distributed systems \cite{tanenbaum2007distributed, cachin2011introduction}, and classical network \cite{kurosecomputer} methods for optimizing reliable quantum computations.

This paper contributes to multiple areas of reliable quantum computer engineering. The entire stack of fault-tolerant quantum computations will be influenced by the data bus, because it changes how high-level circuits are designed, how quantum chip can be layed out, as well as to how distributed quantum architectures are implemented.

\section*{Acknowledgements}
DH is supported by the RIKEN IPA program. AP was supported by project CHARON hosted at the Linz Institute of Technology.

FN is supported in part by the: MURI Center for Dynamic Magneto-Optics via the Air Force Office of Scientific Research (AFOSR) (FA9550-14-1-0040), Army Research Office (ARO) (Grant No. W911NF-18-1-0358), Asian Office of Aerospace Research and Development (AOARD) (Grant No. FA2386-18-1-4045), Japan Science and Technology Agency (JST) (via the Q-LEAP program, the ImPACT program, and the CREST Grant No. JPMJCR1676), Japan Society for the Promotion of Science (JSPS) (JSPS-RFBR Grant No. 17-52-50023, and JSPS-FWO Grant No. VS.059.18N), the RIKEN-AIST Challenge Research Fund, and the John Templeton Foundation.

\section{Appendix}

\subsection{Explicit calculation of mixing different bases}

\begin{figure}
    \centering
    \includegraphics[width=0.4\linewidth]{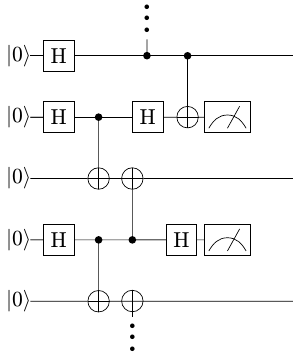}
    \caption{Preparation and syndrome measurement for a mixed GHZ state as described by Equation~\ref{eq:mixed_GHZ}.}
    \label{fig:mixed_GHZ_stabilizers}
\end{figure}

To be able to measure the joint parity of different bases, such as a logical $XZ$-measurement, the data bus has to be prepared in a state of the following form:

\begin{equation}
	\ket{\psi} = \ket{00\ldots 0++\ldots +} + \ket{11\ldots 1--\ldots -}
    \label{eq:mixed_GHZ}
\end{equation}

\begin{figure}[t]
    \centering
    \includegraphics[width=\linewidth]{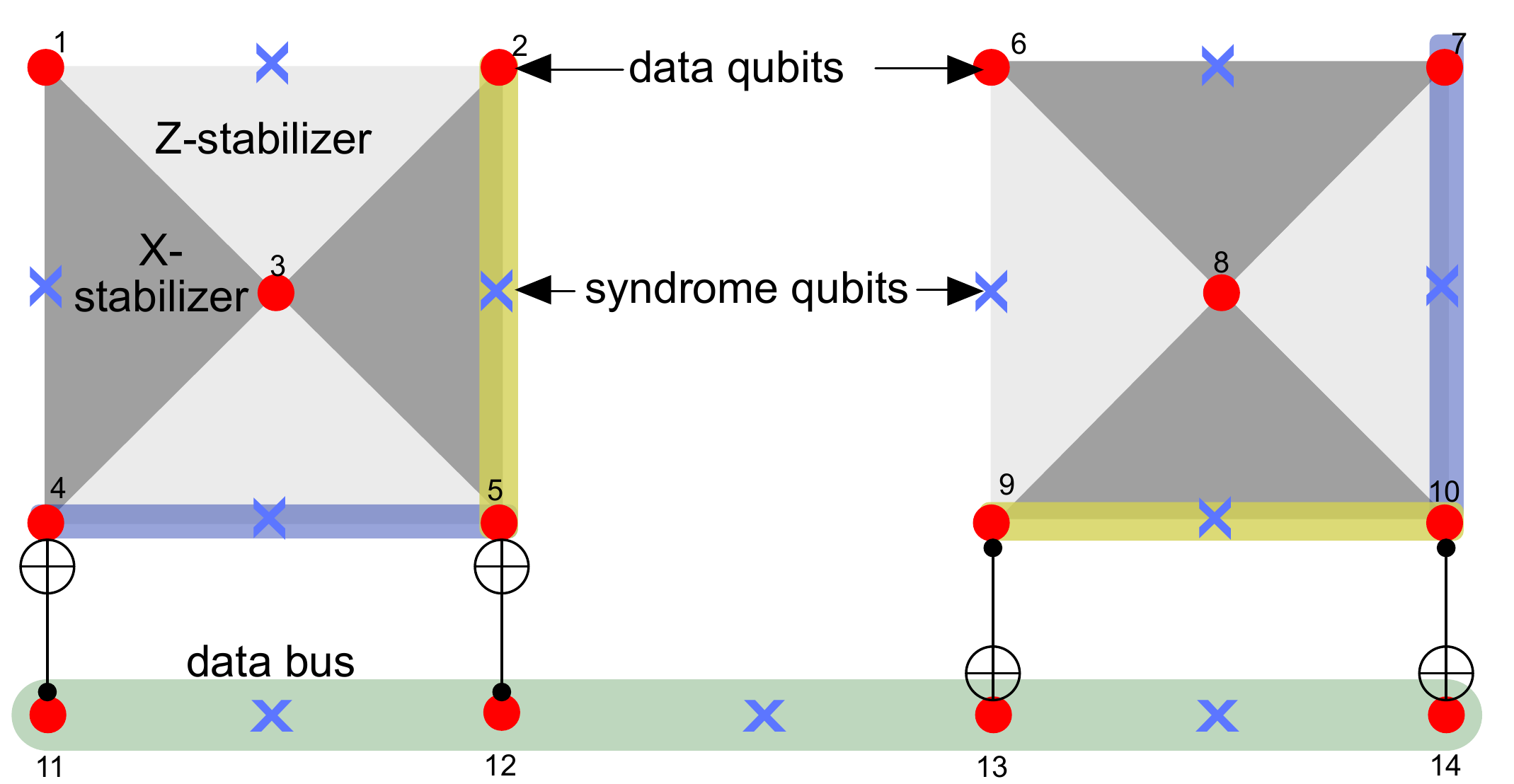}
    \caption{The measurement procedure for two planar code patches in different bases (rotated). The GHZ-state has a basis change from $Z$-basis to the $X$-basis to be able to encode both the logical $Z$-operator (left) and logical $X$-operator (right). The CNOT operations need to be flipped. It can be confirmed that this gives the correct logical operation by calculating the stabilizer table.}
    \label{fig:different_basis_measurement}
\end{figure}

The first part of this GHZ state describes the logical $X$ contribution and the latter part describes the $Z$ contribution of the parity check. This GHZ state can be equally verified using stabilizer measurements shown in~\ref{fig:mixed_GHZ_stabilizers}. In Figure~\ref{fig:different_basis_measurement} the surface code patches are rotated appropriately, to be able to connect different bases of the logical qubits. The state, before the CNOTs are applied, is given by the following stabilizer table:
\begin{equation*}
\begin{smallmatrix}
  1&2&3&4&5&6&7&8&9&10&11&12&13&14 \\
  X&&X&X& &&&& & & & & & \\
   & X&X& &X &&&& & & & & &\\
  Z&Z &Z&& &&&& & & & & &\\
  &&Z&Z &Z &&&& & & & & & \\
  &&&& &Z&&Z&Z & & & & &\\
  &&&& &&Z&Z& &Z & & & &\\
  &&&& &X&X&X& & & & & &\\
  &&&& &&&X&X &X & & & &\\
  &&&& &&&&& &X & X&Z &Z\\
  &&&& &&&&& &Z &Z & &\\
  &&&& &&&& & & & Z&X &\\
  &&&& &&&& & & & &X &X\\
\end{smallmatrix}
\end{equation*}
The first part of this table corresponds to the leftmost planar code patch. The second block consists of the right surface code patch and the last qubits (11-14) in the stabilizer table describe the mixed GHZ state. Now the CNOTs are applied resulting in:
\begin{equation*}
\begin{smallmatrix}
  1&2&3&4&5&6&7&8&9&10&11&12&13&14 \\
  X&&X&X& &&&& & & & & & \\
   & X&X& &X &&&& & & & & &\\
  Z&Z &Z&& &&&& & & & & &\\
  &&Z&Z &Z &&&& & & \textcolor{darkred}{Z}& \textcolor{darkred}{Z}& & \\
  &&&& &Z&&Z&Z & & & & &\\
  &&&& &&Z&Z& &Z & & & &\\
  &&&& &X&X&X& & & & & &\\
  &&&& &&&X&X &X & & &\textcolor{darkred}{X} &\textcolor{darkred}{X}\\
  &&\textcolor{darkred}{X}&\textcolor{darkred}{X}& &&&&\textcolor{darkred}{Z}&\textcolor{darkred}{Z} &X & X&Z &Z\\
  &&&& &&&&& &Z &Z & &\\
  &&&& &&&& & & & Z&X &\\
  &&&& &&&& & & & &X&X\\
\end{smallmatrix}
\end{equation*}
The $XZ$-parity can be deduced using the total measurement result of the mixed GHZ-state. The data bus qubits that are responsible for the $X$ contribution of the logical parity check have to be measured in the $X$ basis. Similarly, the $Z$ contribution has to be measured in the $Z$-basis. From the total parity of the GHZ state one can deduce the result of the $XZ$-measurement.
\subsection{Explicit calculation of the $Y$-state measurement}

In the following, we will explicitly calculate the stabilizers for the smallest patches to see how a $Y$-state measurement can be performed using the data bus. 

\begin{figure}[t]
    \centering
    \includegraphics[width=0.5\linewidth]{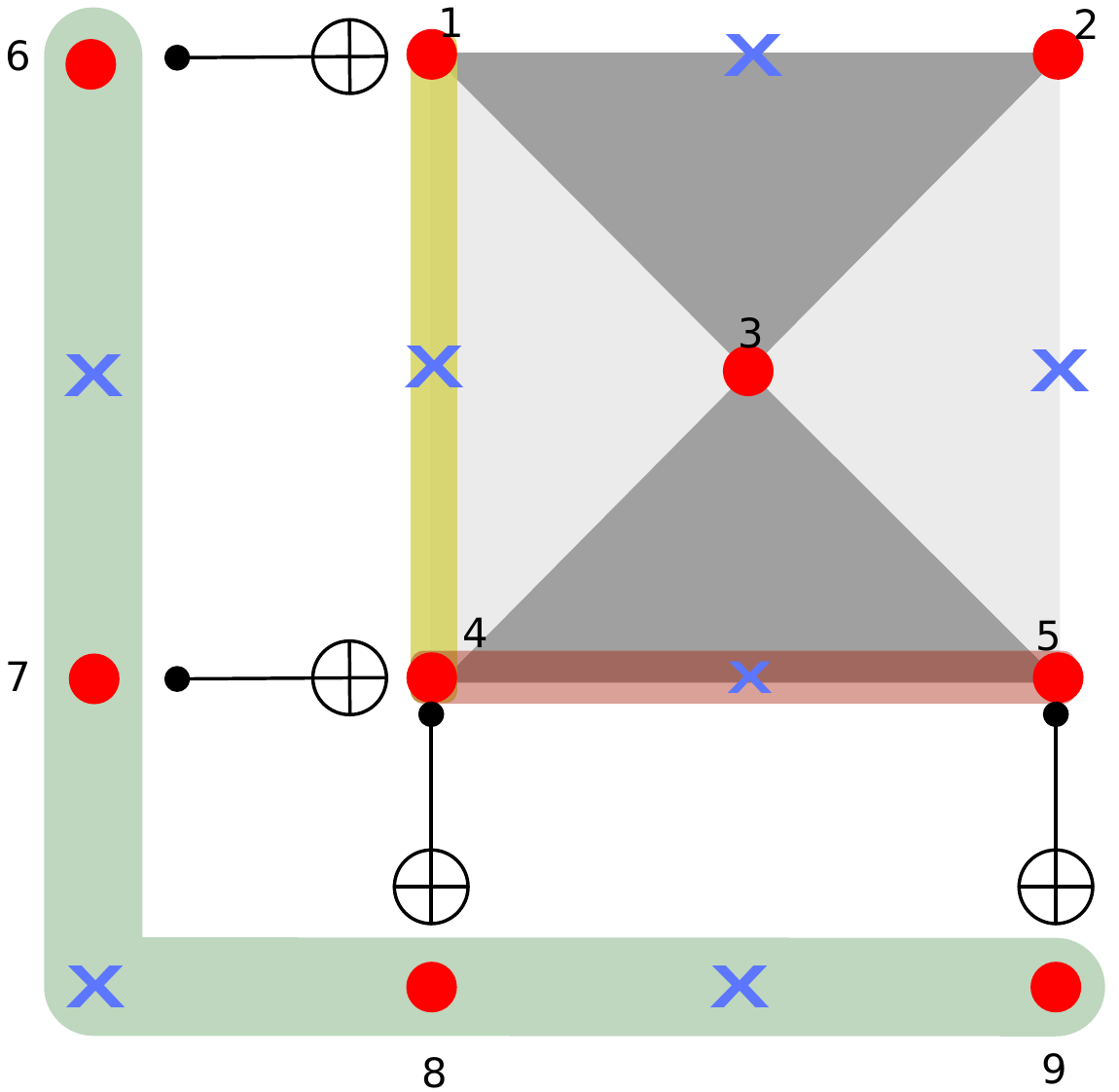}
    \caption{The measurement procedure for the $Y$-state. The operation builds upon the idea of mixing bases in the GHZ-state as was the case before, when patches were rotated. The order in which the transversal CNOTs are applied is important to deduce the $Y$-state from the parity measurement of the GHZ-state.}
    \label{fig:Y-state_measurement}
\end{figure}

Two transversal CNOTs are applied in Figure~\ref{fig:Y-state_measurement}: 1) one for the $X$-state, and 2) another one for the $Z$-state. The total parity stored in the GHZ-state is correlated to the $Y$-state measurement result. These two transversal CNOT operations anti-commute. The order in which the two CNOTs are performed is required to deduce the measurement result from the parity. This will become obvious in the following stabilizer calculation. Before the application of the CNOTs, the state is given by:

\begin{equation*}
\begin{smallmatrix}
  1&2&3&4& 5&6&7&8& 9 \\
  X&X&X&& &&&& \\
   & & X&X&X &&&& \\
  Z& &Z&Z& &&&&\\
  &Z&Z& &Z &&&& \\
  &&&& &X&X&Z&Z\\
  &&&& &Z&Z&& \\
  &&&& &&Z&X& \\
  &&&& &&&X&X \\
\end{smallmatrix}
\end{equation*}

The numbers for each qubit are shown in Figure~\ref{fig:Y-state_measurement}. The stabilizers towards the upper left of the stabilizer table originate from the distance-2 planar-code patch and the stabilizers towards the bottom right stem from the GHZ-state. One can see that the GHZ-state has a basis-change halfway through because the parity of both the logical $X$ and $Z$-operator has to be measured. In order to avoid error propagation, the following schedule for the application of CNOTs~\cite{schedule_error_propagation1,error_rate_wang} should be used: north, south, east, west. Thus, the CNOTs contributing to the $X$-operator are performed first. This changes the stabilizer table to the following one:
  \begin{equation*}
    \begin{smallmatrix}
      1&2&3&4& 5&6&7&8& 9 \\
      X&X&X&& &&&& \\
       & & X&X&X &&&X&X \\
      Z& &Z&Z& &&&&\\
      &Z&Z& &Z &&&& \\
      &&&Z&Z&X&X&Z&Z\\
      &&&& &Z&Z&& \\
      &&&& &&Z&X& \\
      &&&& &&&X&X \\
    \end{smallmatrix}
  \end{equation*}
Now, the CNOTs along the vertical part of the GHZ-state are performed changing the stabilizer table to:
  \begin{equation*}
    \begin{smallmatrix}
      1&2&3&4& 5&6&7&8& 9 \\
      X&X&X&& &&&& \\
       & & X&X&X &&&X&X \\
      Z& &Z&Z& Z&Z&&&\\
      &Z&Z& &Z &&&& \\
      X&&&ZX&Z&X&X&Z&Z\\
      &&&& &Z&Z&& \\
      &&&& &&Z&X& \\
      &&&& &&&X&X \\
    \end{smallmatrix}
  \end{equation*}

The logical $Y$-operator is given by $Y = i XZ = i \left(X_1 X_4\right)\left(Z_4 Z_5\right)$. After measuring the parity of the GHZ-state ($X_6X_7Z_8Z_9$) one can deduce the result of the $Y$ state. However, one first has to permute the $Z_4X_4 = - X_4 Z_4$, and thus flip the parity result to obtain the $Y$-measurement result.

\subsection{Rotated patches and their layout}

\begin{figure}
    \centering
    \includegraphics[width=\linewidth]{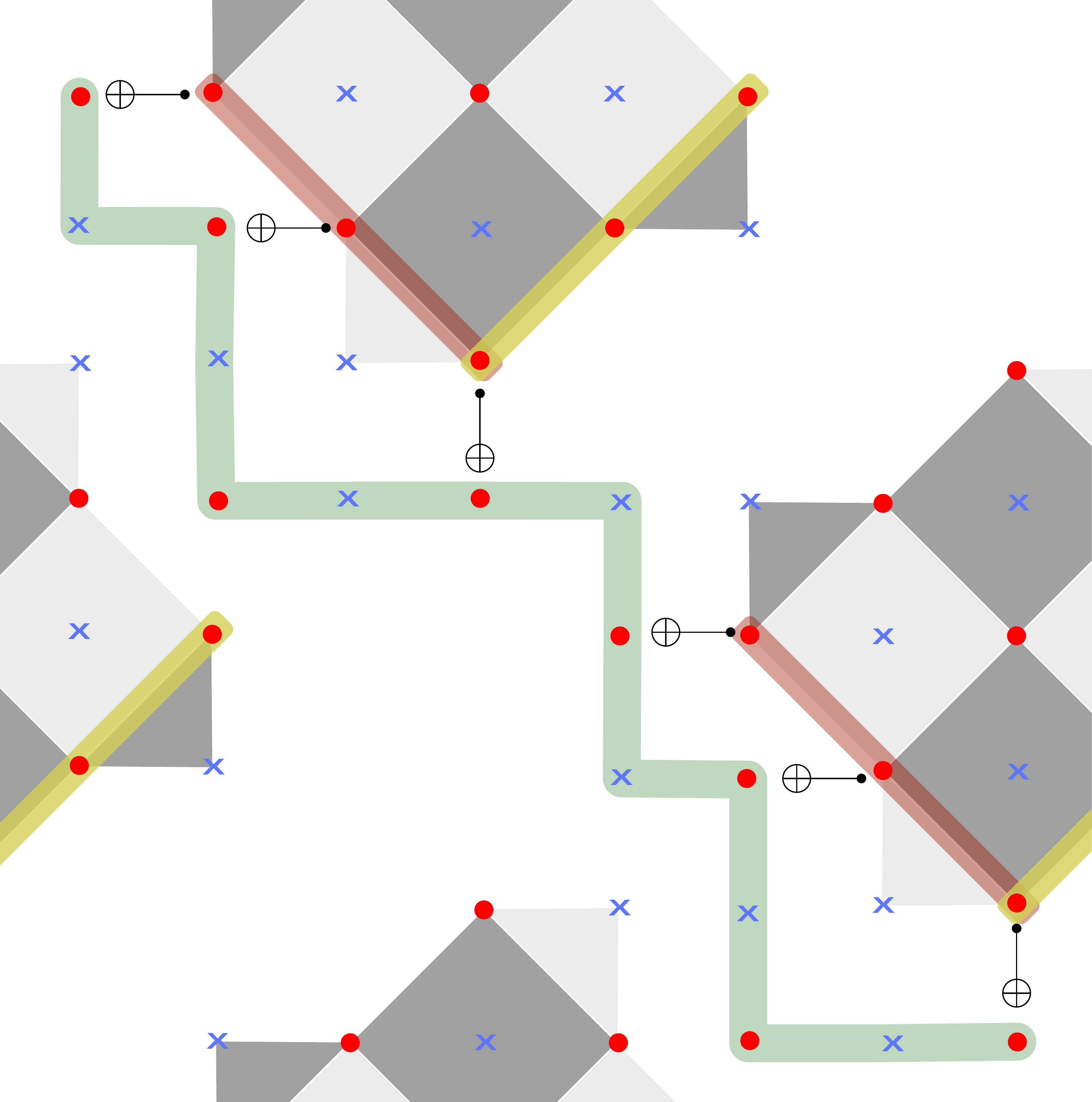}
    \caption{Layout of the data bus between rotated patches: The qubits needed for merges and splits can be used for the GHZ-chain. The GHZ-chain is colored green and each red dot indicates a data qubit. The blue crosses are the data qubits used for syndrome extraction.}
    \label{fig:rotated_patch_layout}
\end{figure}

The formalism developed in this work can also be applied to rotated patches. In Figure~\ref{fig:rotated_patch_layout} the arrangement is shown for four patches. The stabilizer measurements of rotated patches have to be performed around the box of data qubits. This requires the GHZ-state to be on a zig-zag path around the patches. Nevertheless, the data bus qubits are also required for merges and thus no additional qubits are required.

\subsection{Intuition behind the trade-off plots}

\begin{figure}
    \centering
    \includegraphics[width=\linewidth]{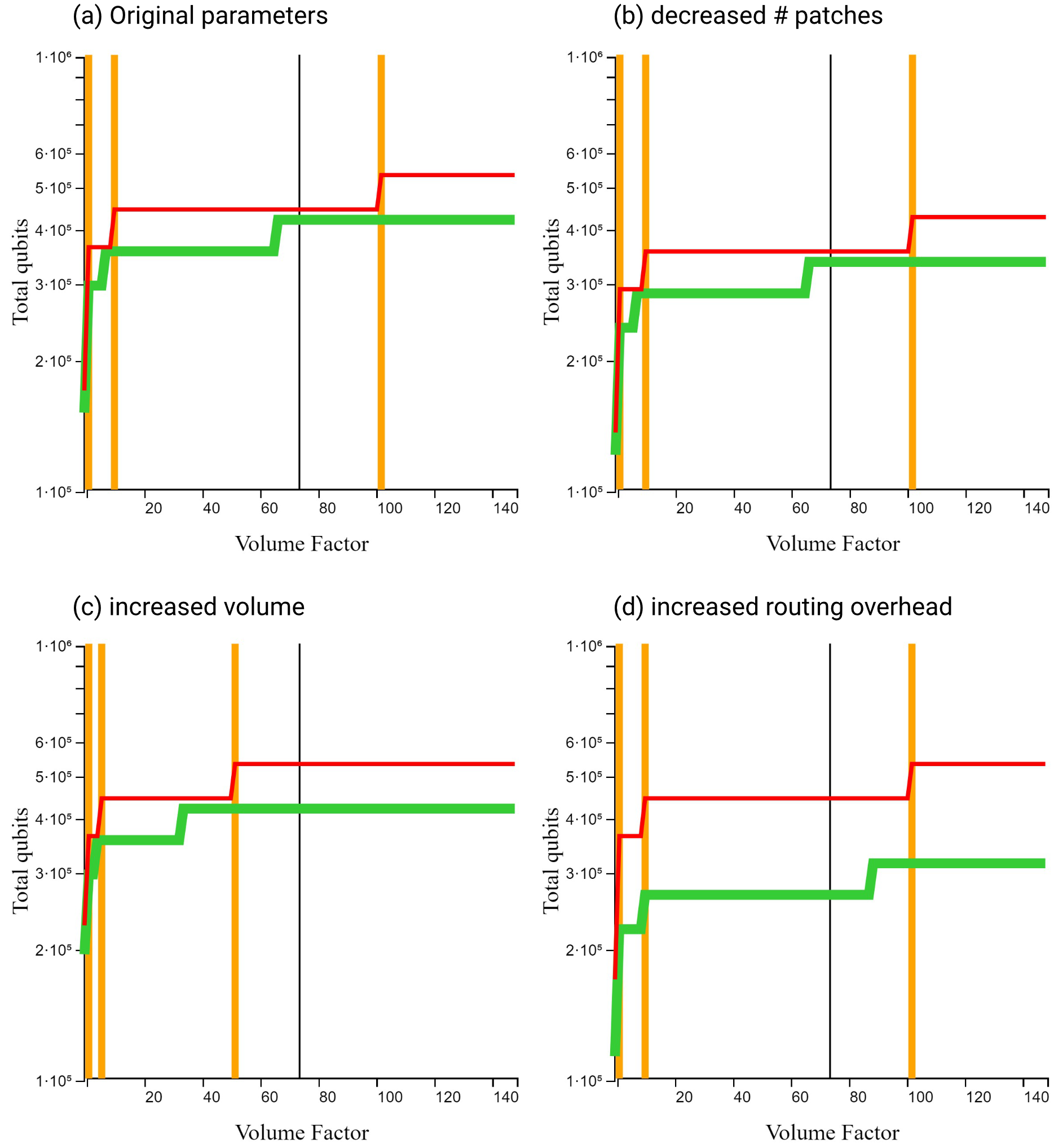}
    \caption{This figure will give some intuition, how a change in parameters affects the performance of the data bus. The initial volume is given by $10^6$ with $500$ surface-code patches and a routing overhead of one third. For a reduction of $100$ (logical) qubit patches, the total number of physical qubits decreases equally for both approaches. An increase in volume by a factor of $5$ results in a shift to the left, and a larger routing overhead (from $1/3$ to $1/2$) increases the difference in total number of qubits between the two methods.}
    \label{fig:value_comparison}
\end{figure}

The usage of the data bus is accompanied by a preliminary analysis using trade-off plots, such as the ones illustrated in Fig.~\ref{fig:value_comparison}. Their usage can be scripted into design automation tools (see Discussion section). Simple rules for interpreting the plot's meaning are necessary.

The most important conclusion drawn from such plots is related to the \emph{distance between the green and the red line}. The further apart, the higher the reduction of the qubit counts. The parameters governing their form are, as mentioned, for example, in Fig.~\ref{fig:value_comparison}: volume, number of patches, physical error rate, and targeted success rate of the overall computation (expressed as volume and patches number). 

The plots include vertical (orange lines), which are boundaries of so-called \emph{distance bins}. Between two consecutive orange lines the distance necessary to protect the computation is the same. A vertical line is marking the \emph{jump from a distance to the next one} (e.g., from $23$ to $25$).

The vertical orange lines refer to the circuit without data bus (red line). Therefore, in the same distance bin, the red line is parallel to the horizontal axis. Once the red line intersects an orange line, its value (qubit count) is higher, because of the larger distance being used. The green lines expose the same behaviour like the red lines, but their value jumps are delayed, because of the qubit count savings: lower volume implies shorter distance. 

The following rules show how the green and red lines behave with the plot parameters:
\begin{itemize}
    \item increasing volume: the distance bin jumps of the green and red line move to the left;
    \item increasing number of patches: green and red line move downwards;
    \item increasing routing overhead: moves only the green line downwards, the red line is kept at position.
\end{itemize}

\subsection{Approximations for the worst case}

The length of the data bus is governed by the total number $Q$ of logical qubit patches (data patches). The maximum total number of qubits necessary for the bus is achieved when the bus replaces all ancillary patches, and touches on two sides each data patch. The bus includes both data and syndrome qubits, and its qubit count is $B = (2 \times Q) \times (2d + 1)$.

The total number of qubits $T$ necessary for implementing a computation with $Q$ data patches and a $B$-sized bus is $T=\left(4\times d^2 \times Q\right) + B$. It is reasonable to assume that a surface patch layout, in the absence of the data bus, requires $A = 0.5\times Q$ ancilla patches, where $0.5$ is the routing factor. This has been the case for both the small and large circuits, and such parameter was also used in the latest chemistry and Shor's algorithm estimations~\cite{}.

In the following we will present the method used within~\cite{link_open_source} to compute the result of the trade-off analysis. This method was used to generate the data from Table~\ref{tab:res}. For the beginning, let $d_a$ be the distance necessary to fault-tolerantly execute a computation with $\left(Q + A\right)$ surface code patches and of volume $V_a$; this is the layout which does not use the data bus. The target now is to determine $d_b$, which is the distance necessary for implementing the same computation, but having now a volume $V_b = V_a \times d_b$.

An additional worst case assumption is that the bus contributes to the failure-estimation of the entire computation with the same number of logical qubits as the number of patches that it replaces. 

The final worst case assumption is that the logical failure rate of $V_b$ having $d_b$ should be strictly less or equal to the failure rate of $V_a$ having $d_a$. This is a strong limitation, because computational rounding errors (for the moment \cite{link_open_source} does not include arbitrary precision floating-point calculations) may result in too large distances, which determine too high qubit counts.

Even under the previous list of strong assumptions the following \emph{iterative approach} generated the numbers presented in the Results section:
\begin{enumerate}
    \item Consider $Q$, a scaled down version by the routing factor of $(Q+A)$.
    \item Let $d_b$ be initialized to $d_a$.
    \item Compute a volume $V_s=0.5 \times V_a \times d_b$. The quantity $V_s$ is the volume that is scaled down by the introduction of the data bus, and which operates on $(Q+A)$ qubits (patches), but is delayed by $d_b$.
    \item Let $d_s$ be the distance to execute $V_s$ with same safety factor as $V_a$.
    \item If $d_s > d_b$, then a distance jump is necessary (see Intuition behind plots), such that $V_s$ is estimated using the next valid $d_b$ value.
    \item Repeat Steps 4, 5 and 6 until $d_s \leq d_b$ and the safety factor for $V_b$ is larger than for the original computation $V_a$
    \item Compute the qubit count for $d_a$ and for $d_b$. If the first (red line in plots) is larger than the latter (green line in plots), a reduction was achieved.
\end{enumerate}


\bibliographystyle{unsrt}
\bibliography{main}

\end{document}